\documentclass[aps, prl, twocolumn, longbibliography]{revtex4-1}
\usepackage{amsmath}
\usepackage{graphicx} 
\usepackage{amssymb}
\usepackage{bm}
\usepackage[dvipsnames]{xcolor}
\usepackage[colorlinks=true, allcolors=RoyalBlue]{hyperref}
\usepackage{orcidlink}

\newcommand{\rmd}{\mathrm{d}}
\newcommand{\mat}[1]{\mathbf{#1}}
\renewcommand{\vec}[1]{\bm{#1}}

\DeclareMathOperator{\Tr}{Tr}
\newcommand{\T}{\mathsf{T}}

\newcommand{\refeq}[1]{Eq.~(\ref{eqn:#1})}
\newcommand{\reffig}[1]{Fig.~\ref{fig:#1}}

\newcommand{\reftab}[1]{Tab.~\ref{tab:#1}}

\begin{document}

\title{Time-frequency Imprints of Extreme Mass-Ratio Inspirals\\in Confusion Gravitational Wave Background}

\author{Lingyuan Ji \orcidlink{0000-0002-3947-7362}}
\affiliation{Department of Physics, 366 Physics North MC 7300, University of California, Berkeley, CA 94720, USA}

\author{Liang Dai \orcidlink{0000-0003-2091-8946}}
\affiliation{Department of Physics, 366 Physics North MC 7300, University of California, Berkeley, CA 94720, USA}

\begin{abstract}
    Detecting individual extreme-mass-ratio inspirals (EMRIs) is a major science goal of future space-based gravitational wave observatories such as Laser Interferometer Space Antenna (LISA) and TianQin. However, matched-filtering can be challenging as waveform templates are required to be accurate over tens of thousands of orbits. We introduce the time-frequency spectrum as an alternative observable that can be exploited to reveal the chirping of EMRIs at the population level. We analytically calculate this spectrum and its correlators for parameterized populations of slowly chirping sources on quasi-circular orbits, assuming a simplified model of the antenna response for a proof of concept. We then exploit this observable to distinguish between Galactic white dwarf binaries and a possible EMRI population, and quantify the precision at which EMRI population parameters can be determined through a Fisher analysis. We explore several scenarios of EMRI populations and find that this new method may allow us to determine EMRI population parameters at an accuracy level of several percent. Since white dwarf binaries have much longer chirping timescales than the EMRIs do, EMRI population properties can still be determined even if their stochastic gravitational wave background has a power spectrum two orders of magnitude weaker than that of the Galactic white dwarf binaries.
\end{abstract}

\maketitle

\paragraph*{Introduction.}  Extreme-mass-ratio inspirals (EMRIs) are stellar-mass compact objects orbiting supermassive black holes (SMBH) at galactic centers. They are important sources for upcoming space-based gravitational-wave (GW) detectors such as Laser Interferometer Space Antenna (LISA) \cite{LISA:2017pwj} and TianQin \cite{TianQin:2015yph}. The study of EMRIs will test theories on the formation and evolution of compact objects around SMBHs \cite{Hopman:2005vr, Aharon:2016kil, Amaro-Seoane:2018gbb, Sari:2019hot, Naoz:2022rru, Naoz:2023hpz, Speri:2022upm, Kritos:2024upo}, providing insight into the dense nuclear star clusters which might have implications on tidal disruption events and quasi-periodic eruptions. Precision measurement of their orbital evolution will enable stringent tests of curved spacetime geometry in the strong-field regime \cite{Barack:2004wc, Barack:2006pq, Gair:2011ym, Gair:2012nm, Yunes:2013dva, Chua:2018yng}.

The detection of individual EMRI signals however will face significant data analysis challenges \cite{Amaro-Seoane:2007osp, MockLISADataChallengeTaskForce:2006sgi, MockLISADataChallengeTaskForce:2007iof, MockLISADataChallengeTaskForce:2009wir}. A fully coherent search using matched filtering, as has been practiced at LIGO/Virgo/KAGRA, would require searching through $\sim 10^{40}$ waveform templates with accurate phases over $\sim 10^5$ orbits \cite{Gair:2004iv}. Alternative sub-optimal methods include semi-coherent searches and time-frequency searches. For the semi-coherent methods, the signal is segmented before matched filtering, compromising signal-to-noise ratios (SNRs) in return for reduced computational costs \cite{Gair:2004iv}. In the time-frequency method, the signal is first reduced to a time-frequency spectrum, in which EMRI-specific features can be identified using inexpensive algorithms. It has been proposed that maximum ``pixels'' or prominent ``tracks'' could be located in the time-frequency spectrum using a simple maximizer or specialized pattern-recognition algorithms \cite{Wen:2005xn, Gair:2005isa, Gair:2007bz, Gair:2008ec}.

These time-frequency methods have been tested with EMRI waveform injected into mock data and recovered \emph{only one at a time}, and can detect typical BH-SMBH systems out to $1$--$3\, \mathrm{Gpc}$ \cite{Wen:2005xn, Gair:2005isa, Gair:2007bz, Gair:2008ec}.  However, the presence of concurrent sources --- Galactic white dwarf binaries (GWDBs), other EMRIs, and any other sources of prominent time-frequency signatures---may greatly degrade the sensitivity for individual signals except for the very loud ones. Moreover, a large fraction of EMRIs will not be individually resolvable \cite{Barack:2004wc}. Together with GWDBs, they collectively contribute to the confusion noise. 

Unresolved EMRIs differ importantly from the unresolved GWDBs in that the former sources chirp at much faster rates due to larger chirp masses. Consequently, the EMRIs leave a unique imprint in the statistical property of the confusion noise.  In this {\it Letter}, we devise a new method that uses (cross-)correlation of time-frequency spectra to extract information about the EMRI population. The method essentially exploits special cases of the four-point correlations of the GW strain that are sensitive to the chirp masses. Focusing on the superimposed GW signal from many incoherent sources, we extract statistical information about the strain beyond what is available from the commonly studied frequency power spectrum. Devised for detecting the collective GW emission from EMRIs, this new method does not require GW waveform models for individual sources with accurate phase information.

\paragraph*{Notation.}  We introduce the time-frequency basis
\begin{equation}
    \psi(f, \tau; t) \equiv W(t - \tau)\,e^{-2\pi\,i\,f\,t},
\end{equation}
where $W(t) \equiv (2\pi T^2)^{-1/2} e^{- t^2 / 2\,T^2}$ is a Gaussian time window of width $T$. The time-frequency representation of a time series $x(t)$ is defined as
\begin{equation}\label{eqn:time-frequency-rep}
    \widehat x(f, \tau) \equiv \int_{-\infty}^{+\infty} x(t)\, \psi(f, \tau; t)\, \rmd t.
\end{equation}
The basis $\psi(f, \tau; t)$ is essentially the Morlet wavelet, which uses a Gaussian window to segment the signal before performing a Fourier transform. We distinguish this from the standard frequency-domain representation of $x(t)$, i.e.\ a simple Fourier transform without the segmentation, $\widetilde x(f) \equiv \int_{-\infty}^{+\infty} x(t)\,e^{-2\pi i f t}\, \rmd t$.  These two representations are related by a convolution
\begin{equation}\label{eqn:tf-f-relation}
    \widehat x(f; \tau) = \int \widetilde x(f - \xi)\,\widetilde W(\xi)\,e^{-2\pi\,i\,\tau\,\xi}\, \rmd \xi,
\end{equation}
where $\widetilde W(\xi) = e^{-2\pi^2\,T^2\,\xi^2}$ is the Fourier transform of the Gaussian window.

We define the time-frequency spectrum $X(f, \tau) \equiv |\widehat x(f, \tau)|^2$. We denote $X(f_a, \tau_a | f_b, \tau_b) \equiv \widehat x^*(f_a, \tau_a)\,\widehat x(f_b, \tau_b)$ for a pair of time-frequency points $(f_a, \tau_a)$ and $(f_b, \tau_b)$, and introduce a shorthand notation $X^{(ab)}$ for it. We also use the notation $\langle \cdots \rangle$ for complete ensemble averaging over source population and noise realizations.

\paragraph*{The Model.}  For a proof of concept, we consider the upcoming LISA mission and adopt technical simplifications that do not fundamentally undermine the methodology: (1) Doppler effect due to heliocentric orbital motion of the spacecraft constellation is neglected; (2) cartwheel motion of the constellation is neglected such that the antenna response is isotropic; (3) only one strain channel is considered. We hence ignore the strain dependence on sky position and polarization but average over orbital inclination. However, we note that our method can be generalized beyond these simplifications, and the analysis can also be applied to other space-based gravitational-wave missions.

We further assume that the EMRIs are on quasi-circular inspirals. The strain time series then depend on an initial frequency $\nu$, an initial phase $\Phi$, the chirp mass $\mathcal M$, and the luminosity distance $D$:
\begin{equation}\label{eqn:t-strain}
    h_j(t) \equiv A_j \cos[2\pi\,(\nu_j\,t + \mu_j\,t^2 / 2) + \Phi_j],
\end{equation}
where $j=1,\,2,\,\cdots,\,N$ for a total of $N$ sources, each characterized by a set of parameters $\vec \phi_j = \{\nu_j, \Phi_j, \mathcal M_j, D_j\}$. The strain amplitude $A_j = A(\nu_j, \mathcal M_j, D_j)$ and the chirp parameter $\mu_j = \mu(\nu_j, \mathcal M_j)$ are given by
\begin{align}
    A(\nu, \mathcal M, D) &\equiv (4/D)\,\left( G\,\mathcal M / c^2 \right)^{5/3}\,\left(\pi \nu/c \right)^{2/3}, \\
    \mu(\nu, \mathcal M) &\equiv (96/5)\,\pi^{8/3}\,\left(G\,\mathcal M/c^3 \right)^{5/3}\,\nu^{11/3}.
\end{align} 
The stationary phase approximation for the frequency-domain waveform is,
\begin{equation}
\label{eqn:f-strain}
    \widetilde h_j(f) \approx ( A_j/2 )\,|\mu_j|^{-1/2}\,e^{-i\pi\,\mu_j\,\tau^2_j(f) + i\,\Phi_j + \frac{i\,\pi}{4} {\rm sgn}(\mu_j) },
\end{equation}
where $\tau_j(f) \equiv (f - \nu_j) / \mu_j$ is the moment when the $j$-th source chirps to have a frequency $f$.  The time-frequency transform has an analytic expression
\begin{equation}\label{eqn:tf-strain}
    \widehat h_j(f, \tau) = ( 2\pi T_j^2 )^{-1/2}\,e^{-(\tau - \tau_j(f))^2/ 2\,T_j^2}\, \widetilde h_j(f),
\end{equation}
where $T_j^2 \equiv T^2 + i / (2\,\pi\,\mu_j)$.  For each source, the parameter set $\vec \phi_j = \{\nu_j, \Phi_j, \mathcal M_j, D_j\}$ is drawn independently from an underlying distribution $P(\vec \phi)$. We shall consider a phenomenological population model in which $P(\vec \phi) = P_\nu(\nu)\,P_\Phi(\Phi)\,P_{\mathcal M}(\mathcal M)\,P_D(D)$, with $P_\Phi(\Phi)=1/(2\pi)$ and
\begin{align}
    P_\nu(\nu) &= (\nu/\nu_0)^{-11/3} / \nu_0, \label{eqn:frequency-distribution} \\
    P_\mathcal M(\mathcal M) &= (\mathcal M\,\sqrt{2\pi}\, \sigma )^{-1}\,e^{-(\ln \mathcal M - \mu)^2/(2\,\sigma^2) }, \label{eqn:chirp-mass-distribution} \\
    P_D(D) &= 3\,D^2/(D_\mathrm{max}^3 - D_\mathrm{min}^3). \label{eqn:luminosity-distance-distribution}
\end{align}
\refeq{frequency-distribution} corresponds to a power-law frequency distribution describing a population steady state following GW-driven orbital tightening. We only consider sources with frequency higher than $\nu_\mathrm{min} = 0.1\, \mathrm{mHz}$, so that the normalization constant is $\nu_0 = 0.144\, \mathrm{mHz}$.  \refeq{chirp-mass-distribution} is a log-normal distribution for the chirp mass with a mean $\mu_{\mathcal M} = \exp(\mu + \sigma^2/2)$ and a variance $\sigma^2_{\mathcal M} = [\exp(\sigma^2) - 1]\exp(2 \mu + \sigma^2)$.  \refeq{luminosity-distance-distribution} corresponds to uniform extragalactic population in Euclidean space in the distance range $D_\mathrm{min} \leq D \leq D_\mathrm{max}$. We refer to $\{N, \sigma_{\mathcal M}, \mu_{\mathcal M}, D_\mathrm{min}, D_\mathrm{max}\}$ as the \emph{population parameters}. For the phase $\Phi$, $P_{\Phi}(\Phi)$ is a uniform distribution on $[0, 2\pi)$. Using the distribution $P(\vec \phi)$, the ensemble average of any quantity $Q(\vec \phi_j)$ of the $j$-th source is computed by
\begin{equation}\label{eqn:source-ensemble-average}
    \langle Q(\vec \phi_j) \rangle = \int Q(\vec \phi) P(\vec \phi) \, \rmd \vec \phi.
\end{equation}
Since this result is independent of $j$, any sum over sources results in a multiplication by $N$.

We further assume that the instrumental noise $n(t)$, uncorrelated with astrophysical sources, is a stationary Gaussian process with a power spectrum $\mathcal N(f)$ taken from Eq.~(1) of Ref.~\cite{Robson:2018ifk}:
\begin{equation}\label{eqn:noise-spectral-density}
    \langle \widetilde n^*(f)\,\widetilde n(f')\rangle = \delta(f - f')\, \mathcal N(f),
\end{equation}

\paragraph*{Time-frequency Spectrum and Correlation.}
We now study the time-frequency spectrum $S(f, \tau)$ of the signal $s(t) \equiv \sum_{j = 1}^N h_j(t) + n(t)$.  Its ensemble average $M(f, \tau) \equiv \langle S(f, \tau) \rangle$ can be calculated (details provided in the Supplemental Material) for sources with independent initial phases
\begin{equation}\label{eqn:tf-spectrum-mean}
    M(f, \tau) = \sum_{j = 1}^N \langle H_j(f, \tau) \rangle + \langle N(f, \tau) \rangle.
\end{equation}
Using \refeq{tf-strain}, we derive an analytic approximation for the source contributions
\begin{equation}
    \langle H_j(f, \tau) \rangle \approx \frac{P_\nu(f)}{8 \sqrt{\pi}\, T}\,\int P_{\mathcal M}(\mathcal M)\,\rmd \mathcal M\, [ A(f, \mathcal M, D_2) ]^2,
\end{equation}
where we define $D_2 \equiv \langle D^{-2} \rangle ^{-1/2} =  [(D_\mathrm{min}^2 + D_\mathrm{min} D_\mathrm{max} + D_\mathrm{max}^2)/3]^{1/2}$.  When integrating out $\nu$, we make the observation that, due to the exponential factor in \refeq{tf-strain}, $H_j(f, \tau)$ is strongly peaked around $f$, with a very narrow width compared to the necessarily broad distribution $P_\nu(\nu)$. Following this observation, two approximations render the $\nu$ integral analytically tractable: (1) the $\nu$ integral essentially picks out $P_\nu(f)$ from the distribution; (2) the $\nu$ dependence through $\mu (\nu, \mathcal M, D)$ can be ignored by replacing the latter with $\mu (f, \mathcal M, D)$. In what follows, we will frequently make similar approximations and write ``$=$'' instead of ``$\approx$'' in the results.  The remaining $\mathcal M$ integral in general has no closed form, so we evaluate it with a 64-point Gauss-Hermite quadrature.

For the noise term $\langle N(f, \tau) \rangle$, we may approximate $\langle N(f, \tau) \rangle = \mathcal N(f)/(2 \sqrt{\pi}\, T)$, under the assumption that frequency features in $\mathcal N(f)$ are broader than $1/(2\pi T)$.

The covariance $C(f_1, \tau_1; f_2, \tau_2) \equiv \langle S(f_1, \tau_1) S(f_2, \tau_2) \rangle - \langle S(f_1, \tau_1) \rangle \langle S(f_2, \tau_2) \rangle$ of the time-frequency spectrum involves a more challenging calculation. The key simplification is that the 4-strain correlation $\langle \widehat h^*_j\,\widehat h_k\,\widehat h^*_l\,\widehat h_m\rangle$ has a phase factor $\langle \exp[i(-\Phi_j + \Phi_k - \Phi_l + \Phi_m)] \rangle$, which will only be non-zero if $j = k \wedge l = m$ or $j = m \wedge l = k$ for independent sources. We derive (details provided in the Supplemental Material)
\begin{widetext}
\begin{multline}\label{eqn:tf-spectrum-variance}
    C(f_1, \tau_1; f_2, \tau_2) = \left|\sum_{j = 1} ^N \left \langle H^{(12)}_j \right \rangle \right|^2 - \sum_{j = 1}^N \left| \left\langle H^{(12)}_j \right\rangle \right|^2 + \sum_{j = 1}^N \left\langle H^{(11)}_j H^{(22)}_j \right\rangle - \sum_{j = 1}^N \left\langle H^{(11)}_j \right \rangle \left \langle H^{(22)}_j \right\rangle\\
    + \left\langle N^{(12)} \right\rangle \sum_{j=1}^N \left\langle H^{(21)}_j \right\rangle + \left\langle N^{(21)} \right\rangle \sum_{j=1}^N \left\langle H^{(12)}_j \right\rangle + \left\langle N^{(11)} N^{(22)} \right\rangle - \left\langle N^{(11)} \right\rangle \left\langle N^{(22)} \right\rangle.
\end{multline}
\end{widetext}
Here $H_j^{(ab)}$ is the shorthand for $H_j(f_a, \tau_a| f_b, \tau_b)$ and similarly for $N^{(ab)}$.  We emphasize again that the expectation value is independent of the subscript $j$.  Following the same strategy to perform the $\nu$ integral as before, we find for the 2-strain terms
\begin{equation}
    \langle H_j^{(ab)} \rangle = \chi^{(ab)}\,\frac{P_\nu(f_{ab})}{8 \sqrt{\pi}\, T} \int P_{\mathcal M}(\mathcal M)\,\rmd \mathcal M\, [ A(f_{ab}, \mathcal M, D_2) ]^2,
\end{equation}
where we define the dimensionless factor
\begin{multline}\label{eqn:chi-ab}
    \chi^{(ab)} = \chi(f_a, \tau_a | f_b, \tau_b) \equiv \exp\left[ - \frac{(\tau_a - \tau_b)^2}{4\,T^2} \right.\\
    \left. \vphantom{\frac{(\tau_a - \tau_b)^2}{4\,T^2}} -\pi^2\, T^2\,(f_a - f_b)^2 + i\pi\,(f_a - f_b)\,(\tau_a + \tau_b)\right]
\end{multline}
and $f_{ab} \equiv (f_a + f_b) / 2$. Similarly, the 2-noise term is given by $\langle N^{(ab)} \rangle = \chi^{(ab)} \mathcal N (f_{ab}) / (2 \sqrt{\pi}\, T)$.  The 4-strain term can also be derived under the same approximation
\begin{multline}
    \langle H_j^{(11)} H_j^{(22)} \rangle = \frac{P_\nu(f_{12})}{32\sqrt{2\pi}\,T} \int P_{\mathcal M}(\mathcal M)\,\rmd \mathcal M \\
    \times [A(f_{12}, \mathcal M, D_4)]^4\, \chi^{(1122)}\left[\mu(f_{12}, \mathcal M)\right],
\end{multline}
where $D_4 \equiv \langle D^{-4}\rangle ^{-1/4} = (D_\mathrm{min}\,D_\mathrm{max}\,D_2^2)^{1/4}$ and
\begin{multline}
\label{eqn:chi-1122}
    \chi^{(1122)}(\mu) = \left(1 + 4\pi^2\,\mu^2\,T^4 \right)^{-1/2} \\
    \times \exp\left\{ - \frac{2 \pi^2\,T^2 }{1 + 4\pi^2\,\mu^2\,T^4} [(f_1 - f_2) - \mu (\tau_1 - \tau_2)]^2 \right\}.
\end{multline}
The 4-noise terms can be expanded using Isserlis' theorem for the Gaussian random noise, so the last two terms in \refeq{tf-spectrum-variance} are simply $\langle N^{(11)} N^{(22)} \rangle - \langle N^{(11)} \rangle \langle N^{(22)} \rangle = |\langle N^{(12)}\rangle|^2$.

We remark on the behavior of the covariance, \refeq{tf-spectrum-variance}. Without chirping, $\mu \to 0$, the covariance would decay exponentially for $|f_1 - f_2| \gtrsim 1/(2\pi T)$ and $|\tau_1 - \tau_2| \gtrsim T$ due to the window $W(t)$ having a width $T$ and a frequency resolution $1/(2\pi T)$. With chirping, \refeq{chi-1122} introduces an additional correlation for $f_1 - f_2 \approx \mu\,(\tau_1 - \tau_2)$, so that two $(f, \tau)$ points along the same chirping ``track'' are correlated. While this piece arising from chirping is formally $O(1/N)$ compared to the dominant part of the covariance in \refeq{tf-spectrum-variance}, it stands out in the covariance when the two $(f, \tau)$ points are sufficiently far apart, providing information about the source chirp masses.  We also note that this feature is \emph{not} contaminated by stationary Gaussian noise $n(t)$, which only contributes to the covariance when the $(f, \tau)$ points are close.

\begin{figure}
    \centering
    \includegraphics[width = \linewidth]{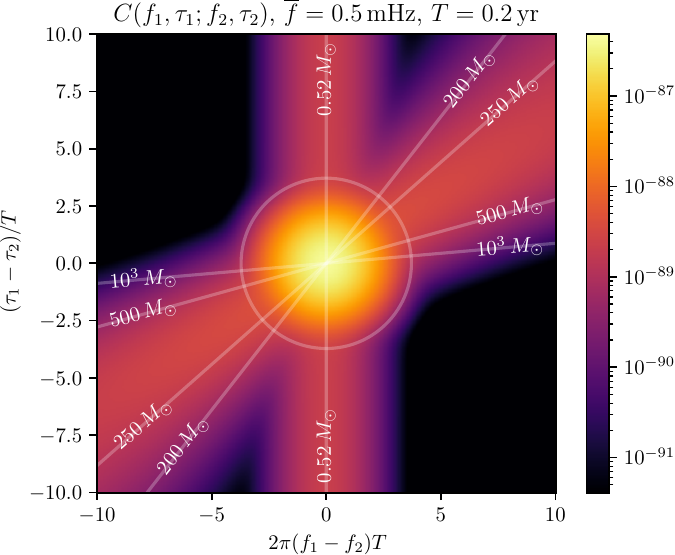}
    \caption{Covariance of the time-frequency spectrum, \refeq{tf-spectrum-variance}, for the Model~I defined in \reftab{model-parameters}, plotted as a function of $(\tau_1 - \tau_2) / T$ and $2\pi (f_1 - f_2) T$.  Here, the center frequency is $\overline f \equiv (f_1 + f_2) / 2 = 0.5\, \mathrm{mHz}$ and the window size is $T = 0.2\, \mathrm{yr}$.  The straight lines indicate the tracks for different chirp masses.  The circle indicates where the 2-strain factor $[\chi^{(12)}]^2$, defined in \refeq{chi-ab}, drops to $10^{-3}$.  Note that the EMRI tracks are not interfering with the GWDB tracks due to the vast difference in chirp mass.  Neither are they degraded by the instrumental noise, which effectively only contributes within the circle.}
    \label{fig:variance}
\end{figure}

The mean, \refeq{tf-spectrum-mean}, and the covariance, \refeq{tf-spectrum-variance}, are derived for sources drawn from a single distribution $P(\vec \phi)$. In reality, both EMRIs and GWDBs exist, so we generalize them to multiple source types described by different population distributions. We assume that each type of sources are described by a distribution of the same parametric form $P(\vec \phi)$ but with different population parameters in \refeq{frequency-distribution}, \refeq{chirp-mass-distribution}, and \refeq{luminosity-distance-distribution}.  Then, we simply extend each sum in \refeq{tf-spectrum-mean} and \refeq{tf-spectrum-variance} to multiple sums, one for each source type. The average in each new sum is then computed for the corresponding source distribution. We plot \refeq{tf-spectrum-variance} in \reffig{variance} for Model~I defined in \reftab{model-parameters}. These models are roughly inline with those in Refs.~\cite{Gair:2017ynp, Babak:2017tow, Benacquista:2006sc}. We emphasize that the abundances of these GW sources, especially of EMRIs, are quite uncertain.

\paragraph*{Likelihood and Fisher Information.} 
We define an array of \emph{observables} $S(f_l, \tau_m)$ at a discrete set of frequencies $f_l$ ($l = 0, 1, \cdots, n_f-1$, uniformly spaced with $\Delta f$ from $f_\mathrm{min}$ to $f_\mathrm{max}$) and window central times $\tau_m$ ($m = 0, 1, \cdots, n_\tau-1$, uniformly spaced with $\Delta \tau$ from $\tau_\mathrm{min}$ to $\tau_\mathrm{max}$). We synthesize the indices $(l,m)$ into one index $k \equiv n_\tau\,l + m$, and arrange the observables in a column vector $\vec d$ of length $n_f \, n_\tau$, such that $d_k \equiv S(f_l, \tau_m)$.  These observables contain information about the population parameters for each type of sources, which we collectively call the \emph{model parameters} $\vec \theta$.  The mean vector $\vec \mu(\vec \theta) \equiv \langle \vec d\rangle$ and the covariance matrix $\mat C(\vec \theta) \equiv \langle \vec d\,\vec d^\T\rangle - \vec \mu\,\vec \mu ^\T$ have been given as functions of $\vec \theta$ in \refeq{tf-spectrum-mean} and \refeq{tf-spectrum-variance}, respectively. We assume a Gaussian likelihood for $\vec \theta$,
\begin{equation} \label{eqn:likelihood}
    \mathcal L(\vec \theta | \vec d) \propto e^{-(1/2) \, [\vec d - \vec \mu(\vec \theta)]^\T \mat C(\vec \theta)^{-1} [\vec d - \vec \mu(\vec \theta)] },
\end{equation}
which is valid in the limit of large source number.

\begin{table}
    \centering
    \begin{tabular}{ccccc}
    \hline\hline
        Parameter \qquad                    & Model I \qquad        & Model II \qquad       & Model III \qquad      & Step size \\\hline
        $N_\mathrm{EMRI}$                   & $\mathbf{3\times10^7}$& $\mathbf{1\times10^7}$& $3\times 10^7$        & $10^{-6}$ \\
        $\mu_{\mathcal M}^\mathrm{EMRI}$    & $250\, M_\odot$       & $250\, M_\odot$       & $250\, M_\odot$       & $10^{-6}$ \\ 
        $\sigma_{\mathcal M}^\mathrm{EMRI}$ & $\mathbf{50\,M_\odot}$& $50\, M_\odot$        & $\mathbf{80\,M_\odot}$& $10^{-6}$ \\
        $D^\mathrm{EMRI}_\mathrm{min}$      & $1\, \mathrm{Mpc}$    & $1\, \mathrm{Mpc}$    & $1\, \mathrm{Mpc}$    & --\footnotemark[1] \\
        $D^\mathrm{EMRI}_\mathrm{max}$      & $10^3\, \mathrm{Mpc}$ & $10^3\, \mathrm{Mpc}$ & $10^3\, \mathrm{Mpc}$ & --\footnotemark[1] \\ \hline
        $N_\mathrm{GWDB}$                   & $3\times 10^7$        & $3\times 10^7$        & $3\times 10^7$        & $10^{-6}$ \\
        $\mu_{\mathcal M}^\mathrm{GWDB}$    & $0.59\, M_\odot$      & $0.59\, M_\odot$      & $0.59\, M_\odot$      & --\footnotemark[1] \\ 
        $\sigma_{\mathcal M}^\mathrm{GWDB}$ & $0.12\, M_\odot$      & $0.12\, M_\odot$      & $0.12\, M_\odot$      & --\footnotemark[1] \\
        $D^\mathrm{GWDB}_\mathrm{min}$      & $8\, \mathrm{kpc}$    & $8\, \mathrm{kpc}$    & $8\, \mathrm{kpc}$    & --\footnotemark[1] \\
        $D^\mathrm{GWDB}_\mathrm{max}$      & $8\, \mathrm{kpc}$    & $8\, \mathrm{kpc}$    & $8\, \mathrm{kpc}$    & --\footnotemark[1] \\ \hline
        $T$                                 & $0.2\, \mathrm{yr}$   & $0.2\, \mathrm{yr}$   & $0.2\, \mathrm{yr}$   & --\footnotemark[1] \\
        $2 \pi T \Delta f$                  & $4.0$                 & $4.0$                 & $4.0$                 & --\footnotemark[1] \\
        $\Delta \tau / T$                   & $4.0$                 & $4.0$                 & $4.0$                 & --\footnotemark[1] \\
        
    \hline\hline
    \end{tabular}
    \newline
    \footnotetext[1]{not included in the information-matrix analysis}
    \caption{Parameters of the Models I, II, and III.  Parameters that are changed across the models are labeled in bold font.  The last column indicates the step sizes used in evaluating the information matrix, \refeq{fisher-information}.}
    \label{tab:model-parameters}
\end{table}

We calculate the Fisher information for source population parameters. The Cram\'{e}r--Rao theorem guarantees that the minimal variance of any non-biased estimator is $\mat F^{-1}$, where $\mat F$ is the Fisher information matrix $F_{\alpha\beta} \equiv - \mathbb E[\partial^2 \ln \mathcal L / (\partial \theta_\alpha \partial \theta_\beta)]$. For the likelihood in \refeq{likelihood}, this becomes
\begin{equation}\label{eqn:fisher-information}
    F_{\alpha \beta} = (1/2)\,\Tr (\mat A_\alpha \mat A_\beta + \mat C^{-1} \mat M_{\alpha\beta}),
\end{equation}
where $\mat A_\alpha \equiv \mat C^{-1} \partial_\alpha \mat C$ and  $\mat M_{\alpha \beta} \equiv (\partial_\alpha \vec \mu) (\partial_\beta \vec \mu)^\T + (\partial_\beta \vec \mu) (\partial_\alpha \vec \mu)^\T$ with $\partial_\alpha \equiv \partial / \partial \theta_\alpha$.  We discuss the numerical techniques necessary to evaluate \refeq{fisher-information} for a large set of observables in the Supplimental Material.

\begin{figure*}
    \centering
    \includegraphics[width = \linewidth]{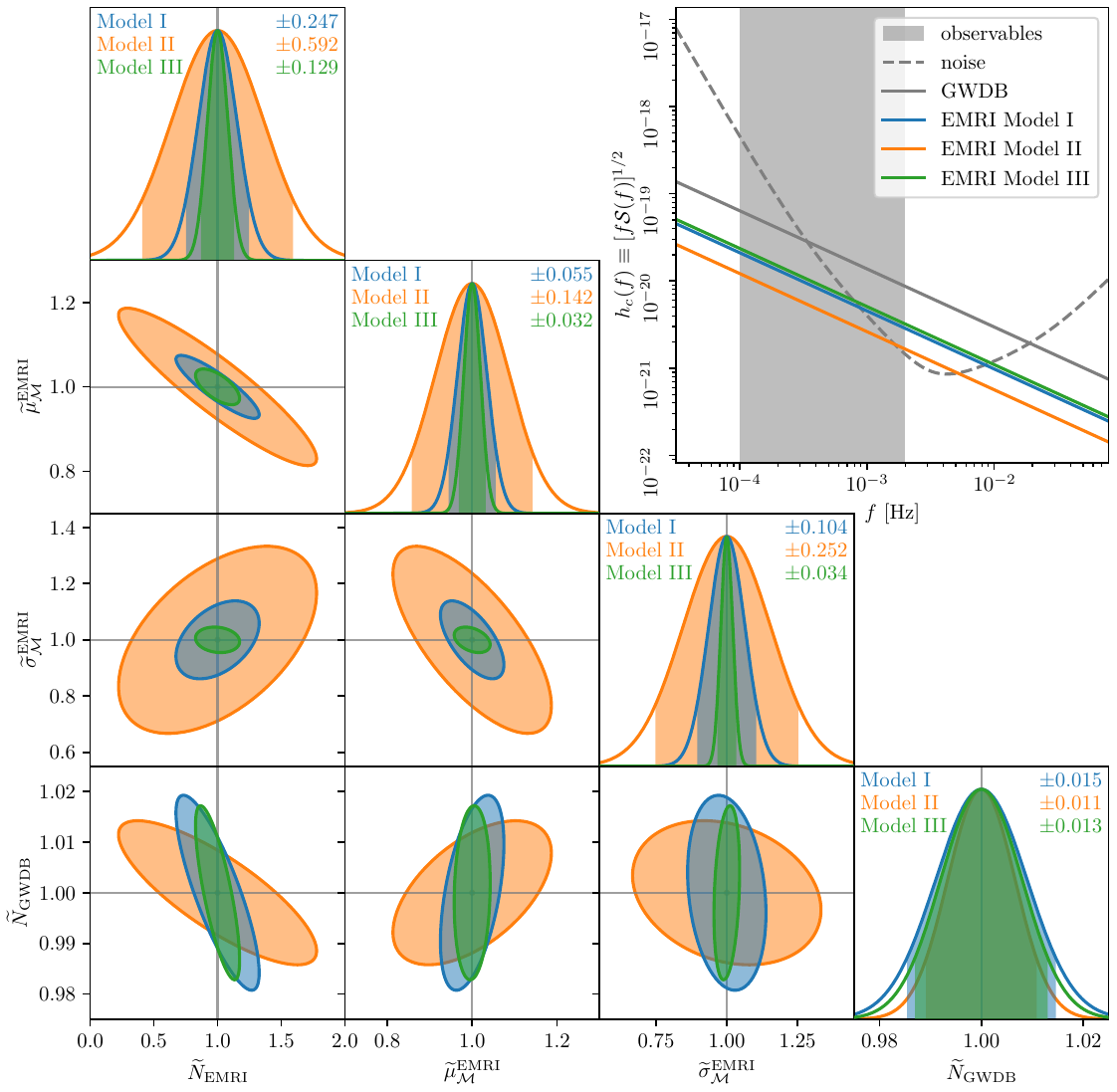}
    \caption{\emph{Lower left triangle plot:}  constrain contours for the central $90\%$ 1D and 2D marginal posterior probability via the information-matrix analysis of the likelihood \refeq{likelihood}.  In the diagonal panels, the fractional uncertainty is labeled for each model.  Note that the constraints are shown for the normalized parameters (symbols with a tilde), so the true value is always unity.  But they correspond to different physical true values listed in \reftab{model-parameters}.  \emph{Upper right plot:}  characteristic strains for each type of sources.  The LISA instrumental noise is plotted with dashed line as a reference.  Note that the EMRI power is lower than the GWDB power in all the models we investigate.  The information comes from the power correlation, i.e.\ the 4-point function of the strain.}
    \label{fig:fisher-results}
\end{figure*}

We investigate three models defined in \reftab{model-parameters} and list the results in \reffig{fisher-results}.  Here, we have chosen $T = 0.2\, \mathrm{yr}$, $[f_\mathrm{min}, f_\mathrm{max}] = [0.1, 2.0]\, \mathrm{mHz}$ with $\Delta f = 4.0 / (2 \pi T) \simeq 1.0 \times 10^{-4}\, \mathrm{mHz}$, and $[\tau_\mathrm{min}, \tau_\mathrm{max}] = [0, 5]\, \mathrm{yr}$ with $\Delta \tau = 4.0\, T = 0.8\, \mathrm{yr}$.  The frequency cutoff, $f_\mathrm{max}$, is chosen such that the shortest gravitational wavelength $c/f_\mathrm{max}$ considered is roughly $1\, \mathrm{AU}$.  This limits the Doppler effect induced by LISA's orbit, which we do not model here.  We demonstrate that this method is capable of determining the EMRI-population parameters to a-few-percent accuracy even when the potential degeneracy from a louder GWDB foreground is allowed.

\paragraph*{Conclusion} We have developed a new method that leverages the time-frequency correlations in the stochastic GW signal to probe a cosmological EMRI population. We have analytically computed this correlation for a set of chirping GW sources under a simplified model of LISA antenna response. We then calculated Fisher information for three different EMRI population models to demonstrate that with this new method one can determine EMRI population parameters to at a-few-percent accuracy, even for an EMRI GW power spectrum that is nearly two orders of magnitude weaker than that expected from the GWDBs.

When LISA data becomes available, \refeq{likelihood} will need to be sampled in order to derive the posterior distribution of the model parameters $\vec \theta$. Direct evaluation of the likelihood would be expensive, costing $\sim 10$ cpu-hours per evaluation in our code. However, the moderate number of dimensions of the population parameter space, $\dim \vec \theta \sim 10$, will allow likelihood evaluation to be emulated with machine-learning techniques such as a Gaussian process or a neural network. Our method is therefore computationally viable for a full Bayesian inference or model selection.

Future work should further address the issue of Doppler frequency shift induced by e.g. LISA's heliocentric orbit, which significantly distorts the GW signal for $f \gtrsim 2\, \mathrm{mHz}$. Future work should also consider the effect of EMRI orbital eccentricity in the time-frequency spectra. An improved parameterization of the EMRI population informed by new astronomical data may reduce the systematic errors of this method.

\paragraph*{Acknowledgment.} The authors would like to thank Nathaniel Leslie, Zack Li, and Wenbin Lu for useful discussions. L.J. acknowledges support from the Berkeley Center for Cosmological Physics (BCCP). L.D. acknowledges research grant support from the Alfred P. Sloan Foundation (Award Number FG-2021-16495), from the Frank and Karen Dabby STEM Fund in the Society of Hellman Fellows, and from the Office of Science, Office of High Energy Physics of the U.S. Department of Energy under Award Number DE-SC-0025293.

\bibliography{references}{}

\onecolumngrid
\appendix

\section{Derivation of \refeq{tf-spectrum-mean} and \refeq{tf-spectrum-variance}}
\label{app:derivation-of-the-covariance}

We outline the derivation of the mean, \refeq{tf-spectrum-mean}, and the variance, \refeq{tf-spectrum-variance}, of the time-frequency spectrum.  We start with \refeq{tf-spectrum-mean},
\begin{align}
    M(f, \tau) &= \left\langle \left[\sum_{j = 1}^N \widehat h_j(f, \tau) + \widehat n(f, \tau)\right]^* \left[\sum_{k = 1}^N \widehat h_k(f, \tau) + \widehat n(f, \tau)\right] \right\rangle  \label{eqn:mean-simplify-1}\\
    & = \sum_{j = 1}^N \sum_{k = 1}^N \left \langle \widehat h_j ^* (f, \tau) \widehat h_k (f, \tau) \right \rangle + \left \langle |\widehat n(f, \tau)|^2\right \rangle \label{eqn:mean-simplify-2}\\
    & = \sum_{j = 1}^N \left \langle H_j(f, \tau) \right \rangle + \left \langle N(f, \tau) \right \rangle \label{eqn:mean-simplify-3}.
\end{align}
Here, in \refeq{mean-simplify-1}, we used the definition of $S(f, \tau)$.  In \refeq{mean-simplify-2}, we used that the instrumental noise and the strains are independent.  In \refeq{mean-simplify-3}, we used that different sources are independent, and then substituted in the shorthand notation $H_j(f, \tau)$ and $N(f, \tau)$.

We now derive \refeq{tf-spectrum-variance}.  We start by calculating the first term,
\begin{align}
    \left\langle S(f_1, \tau_1) S(f_2, \tau_2) \right\rangle 
    &= \left\langle \left[\sum_{j_1 = 1}^N \widehat h_{j_1}^{(1)} + \widehat n^{(1)}\right]^* \left[\sum_{k_1 = 1}^N \widehat h_{k_1}^{(1)} + \widehat n^{(1)}\right] \left[\sum_{j_2 = 1}^N \widehat h_{j_2}^{(2)} + \widehat n^{(2)} \right]^* \left[\sum_{k_2 = 1}^N \widehat h_{k_2}^{(2)} + \widehat n^{(2)}\right] \right\rangle  \label{eqn:variance-simplify-1} \\
    &= \sum_{j_1 j_2 k_1 k_2} \left \langle \left[ \widehat h_{j_1}^{(1)} \right]^* \widehat h_{k_1}^{(1)} \left[ \widehat h_{j_2}^{(2)} \right]^* \widehat h_{k_2}^{(2)} \right \rangle + \left\langle N^{(22)} \right\rangle \sum_{j} \left\langle H^{(11)}_j \right\rangle + \left\langle N^{(11)} \right\rangle \sum_{j} \left\langle H^{(22)}_j \right\rangle \nonumber\\
    &\qquad \qquad + \left\langle N^{(21)} \right\rangle \sum_{j} \left\langle H^{(12)}_j \right\rangle + \left\langle N^{(12)} \right\rangle \sum_{j} \left\langle H^{(21)}_j \right\rangle + \left\langle N^{(11)} N^{(22)} \right\rangle. \label{eqn:variance-simplify-2}
\end{align}
Here, in \refeq{variance-simplify-1}, we again used the definition of $S(f_1, \tau_1)$ and $S(f_2, \tau_2)$.  In \refeq{variance-simplify-2}, we expanded the product and used the statistical independence between the instrumental noise and the sources, and between two different sources.  From here we will also use the shorthand $H_j^{(ab)} = H_j(f_a, \tau_a | f_b, \tau_b)$ and $N^{(ab)} = N(f_a, \tau_a | f_b, \tau_b)$.  We now compute the 4-strain term [the first term in \refeq{variance-simplify-2}] by classifying the sum on indices $(j_1, j_2, k_1, k_2)$ into three cases that are mutually exclusive and collectively exhaustive: 1) $j_1 = k_1 \neq j_2 = k_2$, 2) $j_1 = k_2 \neq j_2 = k_1$, and 3) $j_1 = k_1 = j_2 = k_2$,
\begin{align}
    (\text{4-strain term})
    &= \sum_{j_1 \neq j_2} \left\langle H^{(11)}_{j_1} \right\rangle \left\langle H^{(22)}_{j_2} \right\rangle + \sum_{j_1 \neq j_2} \left\langle H^{(12)}_{j_1} \right\rangle \left\langle H^{(21)}_{j_2} \right\rangle + \sum_j \left\langle H^{(11)}_j H^{(22)}_j\right\rangle \label{eqn:variance-simplify-4}\\
    &= \sum_{j_1 j_2} \left\langle H^{(11)}_{j_1} \right\rangle \left\langle H^{(22)}_{j_2} \right\rangle
    - \sum_j \left\langle H^{(11)}_j \right\rangle \left\langle H^{(22)}_j \right\rangle
    + \left| \sum_j \left\langle H^{(12)}_j \right \rangle \right|^2
    - \sum_j \left| \left\langle H^{(12)}_j \right \rangle \right|^2
    + \sum_j \left\langle H^{(11)}_j H^{(22)}_j\right\rangle. \label{eqn:variance-simplify-5}
\end{align}
Here, in \refeq{variance-simplify-4}, we used the independence between two different sources.  In \refeq{variance-simplify-5}, we modified the sums $\sum_{j_1 \neq j_2}$ with restrictions $j_1 \neq j_2$ into sums $\sum_{j_1 j_2}$ without restrictions and subtracted the difference.  \refeq{tf-spectrum-variance} is then obtained by plugging \refeq{variance-simplify-5} into \refeq{variance-simplify-2} and subtracting $\langle S(f_1, \tau_1) \rangle \langle S(f_2, \tau_2) \rangle$ using \refeq{mean-simplify-3}.

\section{Computing Fisher Information}
\label{app:computing-the-information-matrix}

Brute-force evaluation of \refeq{fisher-information} is computationally infeasible due to the enormous matrix size of the problem. For a uniform frequency spacing $\Delta f$ and a uniform time spacing $\Delta \tau$, the total number of observables is $U \equiv n_f n_\tau \simeq F_\mathrm{obs} T_\mathrm{obs} / (\Delta f \Delta \tau)$. Here $F_\mathrm{obs}$ and $T_\mathrm{obs}$ are the bandwidth and observation time span of LISA, respectively. We expect Fisher information to saturate with spacings $\Delta f \simeq 1/(2\pi T)$ and $\Delta \tau \simeq T$, which gives $U \simeq 10^6$ regardless of $T$. It is prohibitive to load the covariance $\mat C$, which contains $U^2 \simeq 10^{12}$ floating-point numbers, into memory at once.

We solve this problem by noting that the covariance matrix $\mat C$ is the sum of a banded symmetric matrix $\mat B$ and a list of rank-one symmetric matrices $\lambda_i \vec v_i \vec v_i ^\T$ ($i = 1, 2, \cdots, n$),
\begin{equation}
    \mat C = \mat B + \lambda_1 \vec v_1 \vec v_1 ^\T + \lambda_2 \vec v_2 \vec v_2 ^\T + \cdots + \lambda_n \vec v_n \vec v_n ^\T.
\end{equation}
Here, the banded matrix $\mat B$ has the terms proportional to $[\chi^{(12)}]^2$ (i.e.\ the 1st, 2nd, 5th, 6th, 7th, and 8th terms) and the term proportional to $\chi^{(1122)}$ (i.e.\ the 3rd term) in \refeq{tf-spectrum-variance}. This is because the factors $[\chi^{(12)}]^2$ and $\chi^{(1122)}$ vanish for any two frequencies $f_1$ and $f_2$ that are sufficiently different, as no GW source can chirp from one to the other in a time $T_\mathrm{obs}$. The rank-one terms $\lambda_i \vec v_i \vec v_i^\T$ ($i = 1, 2, \cdots, n$) include the $n$ copies of the 4th term in \refeq{tf-spectrum-variance}, each corresponding to a distinct source population (see main text). This is because the 4th term factorizes into a piece that only depends on $(f_1, \tau_1)$ and another piece that only depends on $(f_2, \tau_2)$. This observation leads to the idea that we can store and manipulate the banded matrix $\mat B$ and the rank-one matrices $(\lambda_i, \vec v_i)$ ($i = 1, 2, \cdots, n$) instead of the full dense matrix $\mat C$. 

Over the LISA mission time span, the maximal amount of frequency chirping is $\sim \mu\, T_\mathrm{obs}$, where $\mu$ is the typical chirping parameter. This indicates that $\mat B$ has a matrix bandwidth $u \simeq (\mu T_\mathrm{obs} / \Delta f) n_\tau = \mu T_\mathrm{obs}^2 / (\Delta f \Delta \tau) \simeq 10^3$, and matrix elements beyond the $u$-th sub-diagonals are effectively zero. Therefore, the banded matrix $\mat B$ can be stored in the banded form with a manageable memory size corresponding to $U\times u \simeq 10^9$ floating-point numbers.

The central task in evaluating the Fisher information matrix \refeq{fisher-information} is computing $\mat X \equiv \mat C^{-1} \mat Y$ for a conforming right-hand-side matrix $\mat Y$, which is achievable with the Sherman-Morrison formula, without having to explicitly form the dense matrix. To demonstrate this procedure below, we let
\begin{equation}
    \mat C_k \equiv \mat B + \sum_{i = 1}^k \lambda_i \vec v_i \vec v_i ^\T, \quad \text{so that} \quad \mat C_0 = \mat B \quad \text{and} \quad \mat C_n = \mat C.
\end{equation}
Since $\mat C_k = \mat C_{k-1} + \lambda_k \vec v_k \vec v_k^\T$, using the Sherman-Morrison formula we find a recursion relation for the inverses $\mat C_k^{-1}$
\begin{equation}\label{eqn:inverse-recursion}
    \mat C_k^{-1} = \mat C_{k-1}^{-1} - \frac{\lambda_k \mat C_{k-1}^{-1} \vec v_k \vec v_k^\T \mat C_{k-1}^{-1}}{1 + \lambda_k \vec v_k^\T \mat C_{k-1}^{-1} \vec v_k} = \mat C_{k-1}^{-1} - \frac{\lambda_k \vec w_k \vec w_k^\T}{1 + \lambda_k \vec v_k^\T \vec w_k},
\end{equation}
where, on the right hand side of the second equality, we introduce the auxiliary vector $\vec w_k \equiv \mat C_{k-1}^{-1} \vec v_k$ and use the symmetry of $\mat C_{k-1}$. Expanding the recursion in \refeq{inverse-recursion} and multiplying $\mat Y$, we find
\begin{equation}\label{eqn:solution-recursion-expanded}
        \mat X = \mat C^{-1} \mat Y = \mat B^{-1} \mat Y - \sum_{k=1}^n \frac{\lambda_k \vec w_k \vec w_k^\T \mat Y}{1 + \lambda_k \vec v_k^\T \vec w_k},
\end{equation}
which can be efficiently computed by solving the banded linear problem $\mat B^{-1} \mat Y$ plus some simple additional terms.  We now need a way to compute the auxiliary vectors $\vec w_k$. Multiplying the $k \to k - 1$ version of \refeq{inverse-recursion} to $\vec v_k$, we have
\begin{equation}
    \vec w_k = \mat C_{k-1}^{-1}\,\vec v_k = \mat C_{k-2}^{-1} \vec v_k - \frac{\lambda_{k-1} \vec w_{k-1} \vec w_{k-1}^\T}{1 + \lambda_{k-1} \vec v_{k-1}^\T \vec w_{k-1}}\, \vec v_k.
\end{equation}
Expanding this recursion gives
\begin{equation}\label{eqn:auxiliary-recursion}
    \vec w_k = \mat B^{-1} \vec v_k - \sum_{i = 1}^{k-1} \frac{\lambda_i \vec v_k^\T \vec w_i}{1 + \lambda_i \vec v_i^\T \vec w_i}\,\vec w_i,
\end{equation}
which again can be computed efficiently by solving the banded linear problem $\mat B^{-1} \vec v_k$ ($k=1, 2, \cdots, n$) plus some simple additional terms only involving $\vec w_i$ where $i < k$. To summarize, we first use \refeq{auxiliary-recursion} to build up the auxiliary vectors $\vec w_k$ ($k = 1, 2, \cdots, n$), and then we use \refeq{solution-recursion-expanded} to compute the desired solution $\mat X = \mat C^{-1} \mat Y$. Note that the list of auxiliary vectors do not have to be recomputed for a new right-hand side $\mat Y$.

Another technical difficulty is that evaluating $\Tr(\mat A_\alpha \mat A_\beta)$ requires $U^2 \sim 10^{12}$ linear solves due to the presence of two matrix inverses, which is computationally unrealistic. However, we note the matrix equality $\Tr(\mat A_\alpha \mat A_\beta) = -\Tr[(\partial_\alpha \mat C^{-1})\partial_\beta \mat C]$ allows to reduce the number of matrix inversion to one. After approximating the derivative with mid-point finite difference, the right-hand side only requires $2\,U \simeq 10^6$ linear solves to evaluate, comparable to the cost of evaluating the $\Tr(\mat C^{-1} \mat M_{\alpha\beta})$ term.

Altogether, we evaluate the Fisher information matrix \refeq{fisher-information} as $F_{\alpha\beta} = \Tr \mat K_{\alpha\beta}$, where 
\begin{equation}\label{eqn:K-matrix}
    \mat K_{\alpha\beta} \equiv \frac{1}{8\Delta \theta_\alpha \Delta \theta_\beta} \left\{ -\left[(\mat C^+_\alpha)^{-1} - (\mat C^-_\alpha)^{-1}\right] \left[\mat C^+_\beta - \mat C^-_\beta\right] + \mat C^{-1} \left[(\vec \mu_\alpha^+ - \vec \mu_\alpha^-) (\vec \mu_\beta^+ - \vec \mu_\beta^-)^T +(\vec \mu_\beta^+ - \vec \mu_\beta^-) (\vec \mu_\alpha^+ - \vec \mu_\alpha^-)^T\right]\right\}.
\end{equation}
Here, $\mat C$ is the covariance matrix of the fiducial model, evaluated at parameters $\vec \theta$. The perturbed mean $\vec \mu^\pm_\alpha$ and the perturbed covariance matrices $\mat C^\pm_\alpha$ are evaluated by varying the parameter $\theta_\alpha$ to $\theta_\alpha \pm \Delta \theta_\alpha$, while keeping all other parameters fixed. Columns of $\mat K_{\alpha\beta}$ are evaluated by acting $\mat C^{-1}$ and $(\mat C^\pm_\alpha)^{-1}$ to different right-hand sides as indicated in \refeq{K-matrix} using the method in \refeq{solution-recursion-expanded}. Note that we do not need to explicitly store the whole $\mat K_{\alpha\beta}$ matrices. It suffices to only store and sum the diagonal elements of $\mat K_{\alpha\beta}$.

As a verification, we have compared (for a small and manageable version of the problem) the Fisher information matrix $\mat F$ calculated through the procedure developed in this Appendix, to the result $\mat F_0$ derived from a brute-force calculation using dense matrices. We find good agreement at the level $||\mat F - \mat F_0|| / ||\mat F_0|| \lesssim 10^{-10}$ when setting $\mat B$ to be full (i.e.\ $u = U - 1$), where $||\cdot||$ is the Frobenius matrix norm.

\end{document}